\documentclass[%
 reprint,
 amsmath,amssymb,
 aps,
 prx,
]{revtex4-2}

\usepackage{graphicx}% Include figure files
\usepackage{dcolumn}% Align table columns on decimal point
\usepackage{bm}% bold math
\usepackage{bbm}
\usepackage{amsmath}
\usepackage{pifont}
\usepackage{multirow}
\usepackage{xcolor}
\newcommand{\cmark}{\ding{51}}%
\newcommand{\xmark}{\ding{55}}%
\newcommand{\1}[1]{\mathbbm{1}_{#1}}

\begin{document}

\preprint{APS/123-QED}

\title{Deep learning reveals hidden interactions in complex systems}% Force line breaks with \\

\author{Seungwoong Ha}
\author{Hawoong Jeong}%
\altaffiliation[Also at ]{Center for Complex Systems, Korea Advanced Institute of Science and Technology, Daejeon 34141, Korea}%Lines break automatically or can be forced with \\
\email{hjeong@kaist.edu }
\affiliation{%
  Department of Physics, Korea Advanced Institute of Science and Technology, Daejeon 34141, Korea
}%

\date{\today}% It is always \today, today,

\begin{abstract}
  Rich phenomena from complex systems have long intrigued researchers, and yet modeling system micro-dynamics and inferring the forms of interaction remain challenging for conventional data-driven approaches, being generally established by human scientists. In this study, we propose AgentNet, a model-free data-driven framework consisting of deep neural networks to reveal and analyze the hidden interactions in complex systems from observed data alone. AgentNet utilizes a graph attention network with novel variable-wise attention to model the interaction between individual agents, and employs various encoders and decoders that can be selectively applied to any desired system. Our model successfully captured a wide variety of simulated complex systems, namely cellular automata (discrete), the Vicsek model (continuous), and active Ornstein--Uhlenbeck particles (non-Markovian) in which, notably, AgentNet's visualized attention values coincided with the true interaction strength and exhibited collective behavior that was absent in the training data. A demonstration with empirical data from a flock of birds showed that AgentNet could identify hidden interaction ranges exhibited by real birds, which cannot be detected by conventional velocity correlation analysis. We expect our framework to open a novel path to investigating complex systems and to provide insight into general process-driven modeling.
\end{abstract}

\maketitle

%\tableofcontents
\begin{figure*}%[tbhp]
  \centering
  \includegraphics[width=\linewidth]{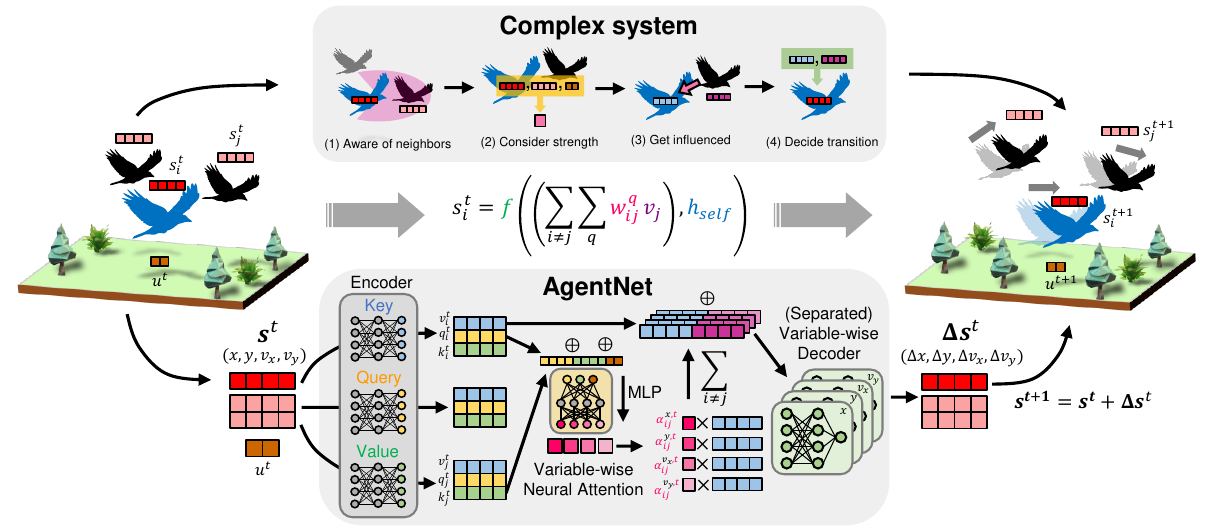}
  \caption{Overview of system formulation and the neural network architecture of the proposed AgentNet. The correspondence between the decision rule of agents in a complex system and a forward pass of AgentNet is depicted. In both panels, the state variable of each agent $s_i^t$ interacts with the state variables of other agents $s_j^t$ in $R_i$ with interaction strength $\alpha_{ij}^t$. The graph attention core learns ${R_i}$ with transformer architecture by encoding $s_i^t$ into key $k_i^t$, query $q_i^t$, and value $v_i^t$, and then calculates the weighted sum of the values of other agents $v_j^t$ according to the variable-wise attention weight $\alpha_{ij}^{q,t}$ as computed by neural attention. Different from GATs, AgentNet assigns attention value for each state variable, and decode it separately to strictly impose the information of variable-wise interaction strength. Other functions, namely $h_{self}$ and $f$, can be captured by both encoder and decoder modules.}
  \label{Fig1} % 여기 수정
\end{figure*}

\section{Introduction}
Complex systems are collections of interactive agents that exhibit non-trivial collective behavior. They have gathered a significant amount of research interest in the last several decades in a wide variety of academic fields from spin systems to human societies. In particular, the domain of physics mainly focuses on investigating the micro-level processes that govern emergent behavior in complex systems and modeling them mathematically. The Vicsek model\cite{VC_origin} is a representative example of such approaches, which attempted to explain collective behaviors of active matter like a bird flock with minimal microscopic description. Unfortunately, due to the intrinsic complexity of these systems, extracting hidden micro-dynamics from the observed data of an unknown complex system is virtually infeasible in most cases. Although conventional process-driven modeling is intelligible and provides the conceptual framework, its application to complex systems, to date, still strongly relies on human intuition with various prior assumptions.

To overcome these obstacles, data-driven modeling (DDM), a methodology that finds a relationship between state variables or their time evolution from observed data, has emerged as a powerful tool for system analysis alongside the emergence of machine learning and large-scale data. In previous literature \cite{DM_automated, DM_distil, DM_unsup4, DM_eq, DM_inverse1, DM_inverse2, DM_tegmark, DM_bacteria, DM_review}, DDM was employed to discover hidden parameters or dynamics from data in an automated manner. Particularly, active matter modeling greatly relies on DDM by first designing a model with intuition from observed data and then performing parameter fitting to match the data \cite{DM_flock1, DM_flock2, DM_flock3, CS_colbehav1, CS_statbird}, although many of them suffer from sparse, noisy, or discontinuous observation data.

Among various DDM techniques, deep neural networks (DNNs) have recently shown phenomenal performance in pattern recognition and function approximation. One specialized DNN variant for graph-structured data is the graph neural network (GNN) \cite{IN_review}, which models dependencies between linked agents on a graph and has enabled remarkable progress in graph analysis. Similar to \cite{IN_original}, one may depict a complex system as a dynamically changing graph in which each vertex is an agent, with links between agents indicating interactions. In this approach, the problem of modeling the micro-dynamics of single agents becomes equivalent to properly inferring the effect from other agents on a graph and estimating the state transition of each agent at the next time step. Several attempts have been separately made to employ GNNs in the prediction and analysis of specific complex systems and physical models \cite{DM_ca, DM_fish, DM_atomic, DM_sociallstm, DM_socialgan, DM_socialattention, IN_interaction1, IN_interaction2, IN_interaction3}, but these approaches are mostly limited to the verification of a single system or a small number of agents, and more significantly, it remains difficult to interpret the characteristics of the interaction due to the neural network's notorious black-box nature. Recently, graph attention networks (GATs) \cite{IN_gat} and its applications showed a path to interpretable GNN by assigning attention to important neighbors, but this attention value cannot be directly interpreted with a physical meaning. For instance, in a multi-dimensional system, the interaction strength cannot be a scalar value since each directional state variable possesses its own interaction range and strength.

Inspired by these recent attempts, we introduce AgentNet, a generalized neural network framework based on GATs with a novel attention scheme to model various complex systems in an physically interpretable manner. AgentNet approximates the transition function of the states of individual agents by training the neural network to predict the future state variables. Due to the rich functional expressibility of DNNs, which is practically unconstrained \cite{IT_uat, IT_exppow}, AgentNet poses minimum prior assumptions about the unknown nature of the target agents. Our model jointly learns the interaction strength that affects \textit{each variable's} transition and overall transition function from observed data in an end-to-end manner without any human intervention or manual operation. This is a critical difference from the conventional approach with GATs, which only assigns a single attention value per agent while our model assigns completely independent attention values for every state variable and employing separate decoders for each of them. We found that our variable-wise attention achieves better performances over GATs, and enables more extensive physical interpretation for the first time that was impossible for conventional GATs such as identifying directional forces separately. Also, the visualization and inspection of the inner modules as granted by our framework enables a clear interpretation of the trained model, which also provides insights for process-driven modeling. As a prediction model, a trained AgentNet can generate an individual level of state predictions from desired initial conditions, making AgentNet an outstanding simulator of target systems including even those that exhibit collective behavior that was absent in the training data.

First, we show the spontaneous correspondence between the complex system and the structure of AgentNet by providing formulations of both systems. The capability of AgentNet is thoroughly demonstrated here via data from simulated complex systems: cellular automata \cite{CA_conway}, the Vicsek model \cite{VC_origin}, and the active Ornstein--Uhlenbeck particle (AOUP) \cite{AOUP_compare} model, along with application to real-world data comprising trajectories in a flock of birds \cite{TC_chimney} containing more than 1800 agents in a single instance, greatly exceeding the previous range of neural network approaches \cite{IN_original, IN_vain, IN_comm} which treated at most several dozens of agents. For the simulated systems, we show that each component of AgentNet learns predictable and tractable parts of the expected transition function by comparing extracted features with ground-truth functions. For the bird flock where the exact analytical expression of the system is completely unknown, AgentNet successfully provides the interaction range of a bird, which is physiologically plausible and coincides with previous behavioral studies about the bird \cite{CS_vf1, CS_vf_xz1}.

\begin{table*}\centering
  \caption{System formulation is applied to simulated model systems. Here, $\mathbf{h} = \sum_j{h_j}$, $\mathbf{h}_{x} = \{ x \text{ directional component of } \mathbf{h} \}$, $\{ \mathbf{h}_{y} = y \text{ directional component of } \mathbf{h} \}$, and $|R_i|$ denote the number of elements of set $R_i$. $r(s_i^t, s_j^t)$ represents the distance between two agents' positions, while $\theta(s_i^t, s_j^t)$ represents the respective angle of the $j$th agent to the $i$th agent. For AOUP, $t+1$ becomes $t+dt$ since the original model is goverened by continuous differential equations.}
  %\centering
  {\renewcommand{\arraystretch}{1.5}
    \begin{tabular}{c|c|c|c}
      \toprule
      System & Cellular automata                                                              & Vicsek model & Active Ornstein--Uhlenbeck Particle \\
      \hline
      $\boldsymbol{s}_i^t$
             & $\{x_i^t, y_i^t, c_i^t\}$
             & $\{x_i^t, y_i^t, v_{xi}^t, v_{yi}^t\}$
             & $\{x_i^t, y_i^t\}$
      \\ \hline
      $u^t$
             & None
             & None
             & $R$
      \\ \hline
      $R_i$
             & $ \{ a_j \in A | r(s_i^t, s_j^t) \leq \sqrt2 \}$
             & $ \{ a_j \in A | r(s_i^t, s_j^t) < r_c, \ |\theta(s_i^t, s_j^t)| < \theta_c\}$
             & $ A - \{ a_i \}$
      \\ \hline
      $w^q$
             & $\begin{aligned}
                &                               \\
          w^c = & \begin{cases}
            1 & \text{if} \ a_j \in R_i    \\
            0 & \text{if} \ a_j \notin R_i
          \end{cases} \\
                &                               \\
        \end{aligned}$
             & $\begin{aligned}
                     &                               \\
          w^x, w^y = & \begin{cases}
            1 & \text{if} \ a_j \in R_i    \\
            0 & \text{if} \ a_j \notin R_i
          \end{cases} \\
                     &                               \\
        \end{aligned}$
             & $\begin{aligned}
          w^x, w^{v_x} \propto & -3(x_i^t-x_j^t)\frac{e^{-r(s_i^t, s_j^t)^{3}/R^{3}}}{R^3} \\
          w^y, w^{v_y} \propto & -3(y_i^t-y_j^t)\frac{e^{-r(s_i^t, s_j^t)^{3}/R^{3}}}{R^3}
        \end{aligned}$
      \\ \hline
      $h_{\text{self}}$
             & $\{ c_i^t \}$
             & $\{ v_{x,i}^t, v_{y,i}^t \}$
             & $\{ x_i^t, y_i^t \}$
      \\ \hline
      $f$
             & $\begin{aligned}

          \Delta x_i^{t+1} & = 0                                                                  \\
          \Delta y_i^{t+1} & = 0                                                                  \\
          \Delta c_i^{t+1} & = \delta_{c_i^t=0}\delta_{\mathbf{h}=3}                              \\
                           & - \delta_{c_i^t=1}(1-\delta_{\mathbf{h}=2})(1-\delta_{\mathbf{h}=3}) \\
        \end{aligned}$
             & $\begin{aligned}

          \\
          \Delta x_i^{t+1}     & = v_{x,i}^{t+1}                                              \\
          \Delta y_i^{t+1}     & = v_{y,i}^{t+1}                                              \\
          \Delta v_{x,i}^{t+1} & = (\mathbf{h}_{x} \ / \ (|R_i| + 1)) + \mathcal{N}(0,\sigma) \\
          \Delta v_{y,i}^{t+1} & = (\mathbf{h}_{y} \ / \ (|R_i| + 1)) + \mathcal{N}(0,\sigma) \\
          \\
        \end{aligned}$
             & $\begin{aligned}
           & \Delta x_i^{t+dt} = v_{x,i}^{t+1}                                                                            \\
           & \Delta y_i^{t+dt} = v_{y,i}^{t+1}                                                                            \\
           & \Delta v_{x,i}^{t+dt} = \mathbf{h}_{x}dt + \frac{\sqrt{2\gamma T}}{\gamma}\mathcal{N}(0,\sqrt{dt}) + f_{x,i} \\
           & \Delta v_{y,i}^{t+dt} = \mathbf{h}_{y}dt + \frac{\sqrt{2\gamma T}}{\gamma}\mathcal{N}(0,\sqrt{dt}) + f_{y,i} \\
           & \Delta f_{x,i}^{t+dt} = -dt/\tau f_{x,i}^{t} + \sqrt{U_0^2 \tau T}\mathcal{N}(0,\sqrt{dt})                   \\
           & \Delta f_{y,i}^{t+dt} = -dt/\tau f_{y,i}^{t} + \sqrt{U_0^2 \tau T}\mathcal{N}(0,\sqrt{dt})                   \\
        \end{aligned}$
      \\ \hline
    \end{tabular}
  } \label{table:1}
\end{table*}

\section{System formulation}

In this paper, we focus on a general agent-based system consisting of $n$ agents for which the state of each agent until time $T$ is (at least partially) identified and observed. The basic premise of the agent-based system is that the agent with the same state variable follows the same decision rule, and the interaction strength between two agents can be fully expressed by their state variable. This implies that any two agents with the same state variables should be interchangeable without altering the outcome.

We denote the set of all $n$ agents as $A = \{ a_1, a_2, \dotsc, a_n \}$ and the corresponding observed state variables of all agents at time $t$ as $\boldsymbol{S}^t = \{ \boldsymbol{s}_1^t, \boldsymbol{s}_2^t, \dotsc, \boldsymbol{s}^t_n \}$, where each state consists of $k$ state variables $\boldsymbol{s}_i^t = \{s_{i,1}^t, s_{i,2}^t, \dotsc, s_{i,k}^t\}$. In addition, the system might have $j$ number of time-dependent global external variables $\boldsymbol{u}^t = \{u_1^t, u_2^t, \dotsc, u_j^t \}$ that affect agent interaction, such as temperature in a thermodynamic system. For simplicity, we abbreviate the set of time series vectors from $t$ to $t-m$, namely $[\boldsymbol{S}^t, \boldsymbol{S}^{t-1}, ... \boldsymbol{S}^{t-m}]$, as $\boldsymbol{S}^{t,m}$.

Generally, agent modeling of a complex system aims to identify the transition function of its constituents through time steps, which can be written as
\begin{align}
  \boldsymbol{S}^{t+1} & = \boldsymbol{S}^{t} + \Delta\boldsymbol{S}^{t+1} \nonumber \\  & = \boldsymbol{S}^{t} + F(\boldsymbol{S}^{t,m}, \boldsymbol{u}^{t,m})  \label{eq:1}
\end{align}
where $m$ is the maximum lag for the system output and $F$ is an overall function that could be deterministic or stochastic. If we focus on the state difference of an individual agent, we can split the overall function $F$ into indvidual transition function $f$ and get
\begin{equation}
  \Delta\boldsymbol{s}_i^{t+1} = f(\boldsymbol{s}_i^{t,m}, \boldsymbol{S}_{\bar{i}}^{t,m}, \boldsymbol{u}^{t,m})\label{eq:2}
\end{equation}

where

\begin{equation}
  \Delta\boldsymbol{S}^{t+1} = [f(\boldsymbol{s}_1^{t,m}, \boldsymbol{S}_{\bar{1}}^{t,m}, \boldsymbol{u}^{t,m}), \dotsc, f(\boldsymbol{s}_n^{t,m}, \boldsymbol{S}_{\bar{n}}^{t,m}, \boldsymbol{u}^{t,m})]\label{eq:3}
\end{equation}

and $\boldsymbol{S}_{\bar{i}}^{t,m}$ indicates that the $i$th agent's state vector $\boldsymbol{s}_i^{t,m}$ is omitted from $\boldsymbol{S}^{t,m}$.

In this study, we assume that the system is mainly dominated by pairwise interactions and higher-order interactions are negligible. Alleviation of this assumption will be discussed in the Conclusion. This means that Eq.~(\ref{eq:2}) becomes

\begin{equation}
  {\Delta}\boldsymbol{s}_i^{t+1} = f(h_{\text{self}}(\boldsymbol{s}_i^{t,m}, \boldsymbol{u}^{t,m}), \sum_{i\neq j} h_{\text{pair}}(\boldsymbol{s}_i^{t,m}, \boldsymbol{s}_j^{t,m}, \boldsymbol{u}^{t,m})) \label{eq:5}
\end{equation}

where $h_{\text{self}}(\boldsymbol{s}_i^{t,m}, \boldsymbol{u}^{t,m})$ denotes self-interaction and $h_{\text{pair}}(\boldsymbol{s}_i^{t,m}, \boldsymbol{s}_j^{t,m}, \boldsymbol{u}^{t,m})$ captures the pairwise interaction between the $i$th and $j$th agents along with the effect of $\boldsymbol{u}^{t,m}$. We note that this generalized formulation encompasses the transition functions of various fundamental systems such as the Monte Carlo simulation of the Ising model \cite{Ising}, the voter model \cite{Voter}, systems governed by Newtonian dynamics, and phase space dynamics driven by the Liouville equation

\begin{table*}[ht!]
  \centering
  \caption{Models to test the performance of AgentNet and their respective characteristics. The $X$s indicate the opposite characteristics: discrete, deterministic, Markovian, and simulated data, respectively.}

  {\renewcommand{\arraystretch}{1.2}
    \setlength{\tabcolsep}{6pt}
    \begin{tabular}{lcccccc}
      \toprule
      System              & Continuity & Stochasticity & Memory effect & Empirical data & Interaction & Remarks              \\ \hline
      Cellular automata   & \xmark     & \xmark        & \xmark        & \xmark         & Discrete    & -                    \\
      Vicsek model        & \cmark     & \cmark        & \xmark        & \xmark         & Discrete    & Generalization       \\
      Active OU particle  & \cmark     & \cmark        & \cmark        & \xmark         & Continuous  & Collective phenomena \\
      Chimney swift flock & \cmark     & \cmark        & \cmark        & \cmark         & Unknown     & Missing data         \\ \hline
    \end{tabular} \label{table:2}
  }
\end{table*}

Although Eq.~(\ref{eq:5}) sums up the interaction with every agent except itself, not every agent is relevant to the transition function of a single agent in a general case. Every agent $a_i$ may have its own interaction range $R_{i} = \{ a_j \in A\ |\ a_i \ \text{interacts with} \ a_j \}$ that can change depending on the current state of the agent, and only a subset (or possibly the entire set) of agents belonging to $R_i$ should be considered. Furthermore, each state variable might be affected by different interaction strengths, e.g. exerting force $F_x$ and $F_y$ can be generally different. Hence, we define the variable-wise interaction strength function between two agents as $w_{ij}^q(\boldsymbol{s}_i^{t,m} \boldsymbol{s}_j^{t,m}, \boldsymbol{u}^{t,m}) \geq 0$ that outputs the $q$-th state variable's interaction magnitude of the $i$th agent, induced by the $j$th agent. Now, $w_{ij}^q$ can be separated from the pairwise interaction function $h(\boldsymbol{s}_i^{t,m}, \boldsymbol{s}_j^{t,m}, \boldsymbol{u}^{t,m})$ to explicitly indicate the variable-wise interaction strength between agents, as follows:

\begin{equation}
  h_{\text{pair}}(\boldsymbol{s}_i^{t,m}, \boldsymbol{s}_j^{t,m}, \boldsymbol{u}^{t,m}) = \sum_q w_{ij}^q(\boldsymbol{s}_i^{t,m}, \boldsymbol{s}_j^{t,m}, \boldsymbol{u}^{t,m}) \boldsymbol{v}_{j}^{t}(\boldsymbol{s}_j^{t,m}). \label{eq:6}
\end{equation}

Note that leftover function $v_j$ conveys information solely from the $j$th agent without loss of generality. So far, we have decomposed an individual transition function into four parts; variable-wise interaction strength function $w^q$, leftover function $v$, self-interaction function $h_{\text{self}}$, and overall function $f$. We note that our formulation aptly applies to a physical system governed by force dynamics by interpreting $w^q$ as the magnitude of a component of an exerting force vector, while the leftover function vector $v$ contains directional information.

In most cases, the exact analytic forms of all these functions ($w^q$, $v$, $h_{\text{self}}$, $f$) are completely unknown, and it is infeasible to elicit these functions from observed data alone. Especially, blindness to variable-wise interaction strength function $w^q$ significantly complicates this inverse problem since we have to test every possible combination of neighbor candidates while simultaneously guessing the correct \textit{nonlinear} functional form of $v$, $g$, and $f$. The problem becomes harder if the system has time-correlation because it expands the range of possibly correlated variable pairs further out in the time dimension. To sum up, many of the current methodologies are not capable of DDM for complex systems without strong prior assumptions regarding the functional form. The proposed framework, AgentNet, successfully tackles this conundrum by employing DNNs to jointly learn all of the aforementioned functions by constructing corresponding neural modules for each of the functions and backpropagating errors from state predictions.

Our formulation of agent-based complex systems is shown in Fig. \ref{Fig1} with corresponding modules in AgentNet: the value vector of the transformer captures the self-interaction $h_{\text{self}}$ and leftover information $v$; variable-wise attention weight $\alpha^q$ captures the interaction magnitude $w^q$; and the weighted sum along with the decoder corresponds to the overall function $f$. This formulation can express all of the model systems used in this study, as described in Table \ref{table:1}.

\begin{figure*}[ht!]
  \centering
  \includegraphics[width=\textwidth]{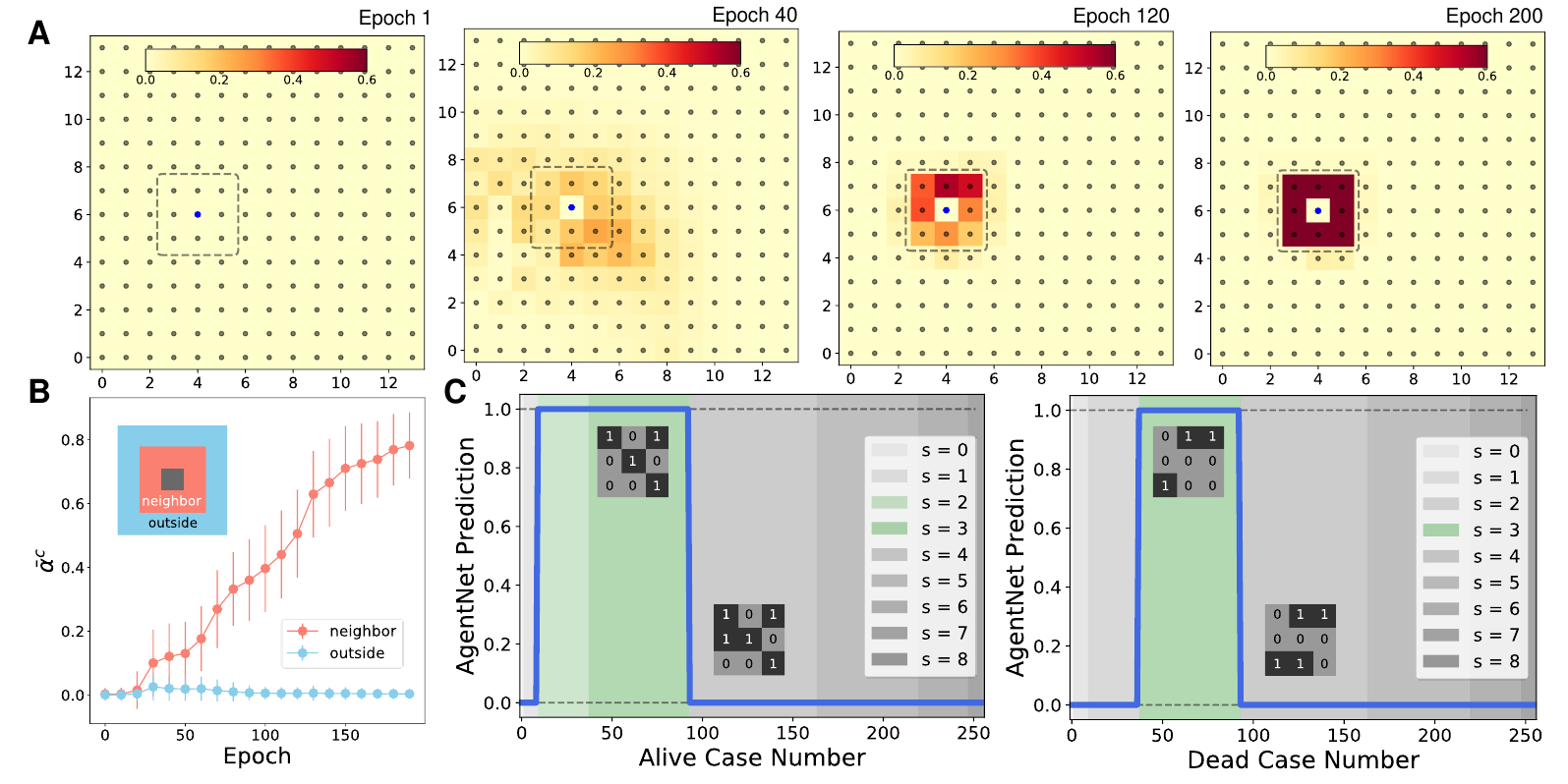}
  \caption{Result of AgentNet for cellular automata. (A) Attention weight transition of a single target cell throughout the training. In the initial stage, the model has no information about the interaction range and assigns near-zero values to all of the cells in the system. Attention gets narrowed down to a smaller region as training advances, and finally concentrates on eight surrounding cells, which is the theoretical interaction range. (B) Attention weight $\bar{\alpha}^c$ of neighbors and outside cells during 200 epochs of training. The attention weight of neighbor cells increases as training proceeds, while the weight of other cells remains $0$. Data is averaged from 100 test samples. (C) AgentNet with respect to given alive (left) and dead microstates (right). The total number of alive cells in the neighborhood is denoted by $s$, which is the sole parameter of the CA decision rule.}
  \label{Fig2}
\end{figure*}

\section{Models and Methods}

AgentNet is a generalized framework for the data-driven modeling of agent-based complex systems, covering most previous works and reinforced with several modifications. The base module of AgentNet, a graph attention module, is similar to a GAT \cite{IN_gat, IN_attention} with transformer architecture \cite{AN_trans}, where each agent decides its next state by putting information from itself and the attention-weighted sum of other agents together. AgentNet initially operates on a fully-connected graph, implying that it initially assumes every agent as a possible neighbor and gradually learns the true interaction partners and strength through training. Our model first encodes the state variables of agent $s^t$ with an encoder, then passes the information to the transformer which computes the impact from the entire system state $\boldsymbol{s^t}$, and finally decodes the outcome with a decoder to obtain the state difference.

In most cases, complex systems have diverse characteristics that are difficult to incorporate into a single modeling framework. As a universal framework, AgentNet resolves this diversity by modifying the encoder and decoder and setting a proper optimization function to fit particular system characteristics while maintaining the core module of the network. In this way, AgentNet addresses a variety of system characteristics such as continuity of state variables, stochasticity of transition function, and memory effects.

First, AgentNet can handle various types of state variables by minimizing cross-entropy for discrete variables and the mean squared error for continuous variables. Second, when the decision rule of a target system is stochastic, there are several ways to construct a neural network with probabilistic output \cite{TC_gan, TC_vae, TC_vrnn, TC_mgn}. AgentNet employs a Gaussian neural network \cite{TC_vrnn} as the decoder of the stochastic AgentNet, which produces means and variances of multiple univariate Gaussian distributions. Lastly, some of the collective phenomena in complex systems appear in non-Markovian settings where past states affect the future state. In this study, we use long short-term memory (LSTM) models for the encoder and decoder of AgentNet to capture (potential) memory effects in the system.

The graph attention module in AgentNet explicitly assigns variable-wise importance $\alpha_{ij}^q$ by first constructing the attention coefficient $\alpha_{ij}^q$ from encoded data $e(\boldsymbol{s^t})$ and applying the sigmoid function to normalize the scale (see Appendix A). We note that the choice of the sigmoid function is crucial because unlike most previous literature \cite{DM_socialattention, IN_attention, IN_gat} where a softmax normalization between agents was used ($\alpha_{ij} = \frac{a_{ij}}{\sum_k{\exp(a_{ik}})}$), here we aim to infer the absolute variable-wise interaction magnitude without normalization among the agents. Also, further differing from conventional approaches for attention coefficients such as additive \cite{AN_additive} and multiplicative \cite{IN_attention} mechanisms, attention coefficients in AgentNet are calculated by multi-layer perceptrons (MLPs) ($\text{Att}$), which enables much more flexible representations (See Appendix C and Fig. \ref{Fig7} for the advantages of neural attention). By virtue of variable-wise separated decoder, attention weights $\alpha_{ij}^q$ \textit{only affects to} $q$-th variable, thus one can identify interaction strengths for each variable by visualizing predicted attention weights. Note that this is different from widely-known multi-headed attention since it feeds concatenated output into a single decoder while AgentNet does not concatenate the output and strictly separates each decoder in order to impose a variable-wise transition function for each attention value, not a mixed overall transition function. In short, AgentNet clarifies the attention scheme from an unknown function of interaction strengths to physically interpretable variable-wise strength. Our study is the first in-depth demonstration of the capability of this form of graph attention scheme, achieved by comparing the attention weights for each variable to the ground-truth interaction strengths in various simulated complex systems.

\begin{figure*}
  \includegraphics[width=\linewidth]{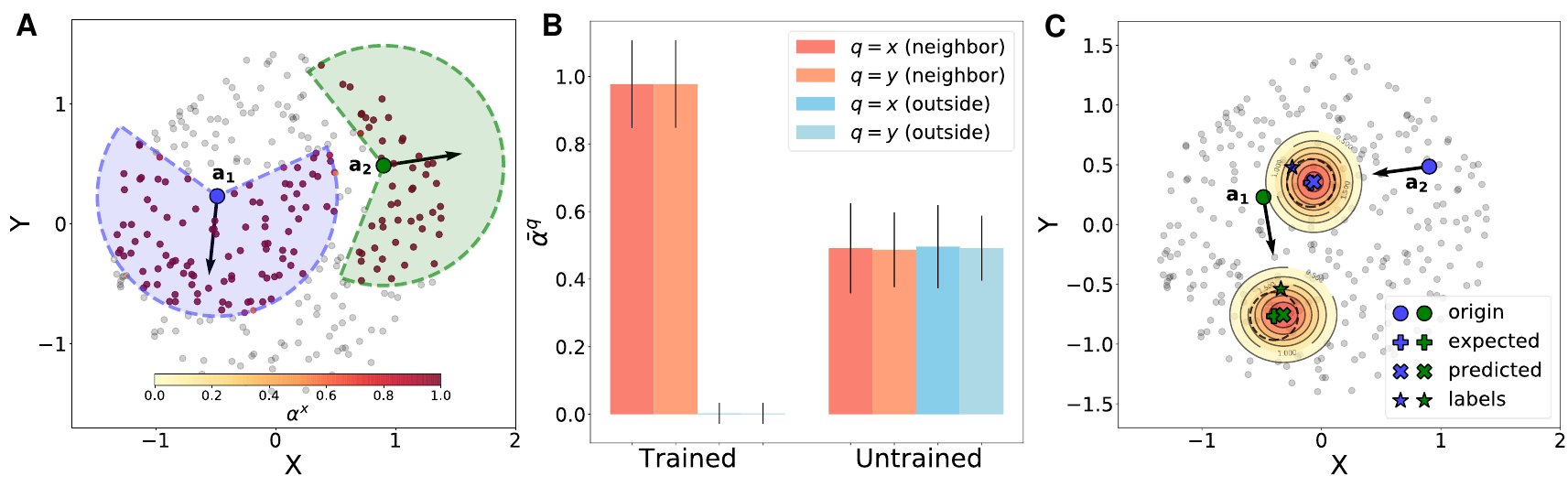}
  \caption{Result of AgentNet for the Vicsek model. Initially, agents are randomly distributed in a circular region with radius $R=\sqrt5$, without any boundary condition. (A) Attention weight visualization of two sample cells, $a_1$ and $a_2$. Both cases show a circular sector of attention distribution with clear boundaries that perfectly matches with ground-truth interaction range. (B) Averaged attention weight for variables $x$ and $y$ before and after training. The fully trained AgentNet learned to identify the neighbor agents and ignore the irrelevant others by assigning near-one and near-zero attention weights, respectively. (C) Position predictions by AgentNet for the two sample cells $a_1$ and $a_2$. Circles indicate the starting positions of the two particles, with the two heatmaps showing the AgentNet prediction along with the means of predicted distributions (Xs). The model predicts the expected theoretical distribution (crosses) with great precision, even when the given training samples (green and blue stars) are distant from the means of the theoretical distribution.}
  \label{Fig3}
\end{figure*}

\section{Results}

This study utilizes three representative complex systems to demonstrate the capacity of AgentNet, along with one empirical dataset for framework evaluation. Table \ref{table:2} summarizes the characteristics of the model systems with an escalating level of complexity. All of the code for model training  and system simulation has been deposited in \cite{github} (see Appendix A for models and baselines implementations, and Appendix B for detailed dataset descriptions and the optimization functions for AgentNet training).

\subsection{Cellular Automata}
First, we verify AgentNet with an older yet fundamental system with rich phenomena, the cellular automata (CA) model. In the CA model, each cell has its own discrete state, either \textit{alive} or \textit{dead}. Each cell interacts with its eight adjacent neighbors, and the state of each cell evolves according to the following two rules. First, a live cell stays alive if two or three neighbor cells are alive. Secondly, a dead cell becomes alive if exactly three neighbor cells are alive. Thus, the interaction strength of CA can be expressed as an indicator function $\mathbbm{1}_{R_i}$ where its value is $1$ if $a_j \in R_i$ and $0$ otherwise.

We simulate CA data in the form of a 14 $\times$ 14 grid of cells with initially randomized states, and the state of the grid after a single time step becomes the target label for each data. AgentNet for CA receives three state variables from each cell: positions $\mathbf{x}^t$ and $\mathbf{y}^t$, and cell state $\mathbf{c}^t$. The output here is a list of expected probabilities that each cell becomes alive. We use the binary cross-entropy loss function between the AgentNet output and the ground-truth label.

Figure \ref{Fig2} summarizes the results of AgentNet for CA, depicting the cell state attention weight $\alpha^c$ of the target cell (in this case, the 102nd cell) across the entire grid. After 120 epochs, AgentNet quickly \textit{realized} that a vast majority of the cells are irrelevant to the target cell, and thereafter concentrated its attention to a more compact region; Fig \ref{Fig2}B shows that AgentNet gradually learns to focus on neighbor cells only. AgentNet was able to figure out the true interaction range after 200 epochs. The result of the prediction test for unseen cases showed perfect 100\% accuracy, as depicted in Fig. \ref{Fig2}C.

\begin{figure*}
  \includegraphics[width=\linewidth]{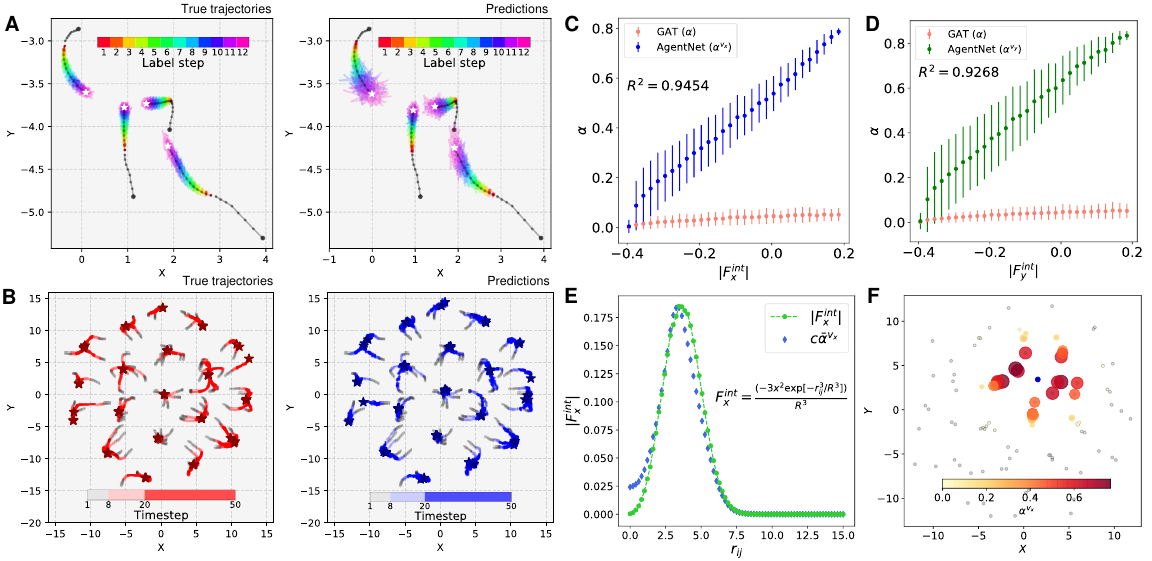}
  \caption{Result of AgentNet for AOUP. (A) In both panels, eight steps of the test data ($R = 4$) of four particles are drawn with black dots starting from the large black dots. AgentNet predictions of the trajectories in the following 12 steps (right panel) perfectly coincide with the sample trajectories from the true Langevin equation (left panel). 100 samples are drawn in both panels, and the final positions are highlighted with white stars. (B) Equilibrium state of a system with $R = 5$, which is unseen at the training stage. A single realization from the true distribution is drawn with final positions marked by red stars (left), while a single sample from the predicted distribution of AgentNet is drawn with final positions marked by blue stars (right). AgentNet for AOUP captures the generalized effect of interaction length $R$ and predicts the collective behavior of the untrained system. (C) Exerted $x$-directional force $F^{\text{int}, x}$ and $x$-directional velocity attention $\bar{\alpha}^{v_x}$ shows a strong linear relationship, while single attention value from GAT does not captures any of the force component. Same holds for (D), for the case of $y$-direction. (E) By plotting relative distance $r_{ij}$ versus force and attention, scaled attention shows good coincidence with the force value up to constant factor $c = 0.28$. (F) Visualization of $\bar{\alpha}^{v_x}$ for a single target particle (blue). Attention and force values for (C) to (E) are collected from 100 test samples with $R = 4$.}
  \label{Fig4}
\end{figure*}

\subsection{Vicsek model}
Next, we validate the capability of AgentNet for a continuous and stochastic system. The Vicsek model \cite{VC_origin} (VM) is one of the earliest and most prominent models to describe an active matter system, where each agent averages the velocity of nearby agents (including itself) to replace its previous velocity. At each time step, every agent updates its position by adding this newly assigned velocity with stochastic noise. In this study, every 300th agent in the simulation interacts with other agents within the range $r_c = 1\ m$ and viewing angle $\theta_c = 120^{\circ}$ of its heading direction. This complex interaction range models the limitations of sight range and angle in real organisms such as birds.

The model receives four state variables, positions $\mathbf{x}^t$ and $\mathbf{y}^t$, and velocities $\mathbf{v_x}^t$ and $\mathbf{v_y}^t$, and predicts the positions of the next time step $\mathbf{x}^{t+1}$ and $\mathbf{y}^{t+1}$ in the form of two one-dimensional (1D) Gaussian distributions by optimizing the sum of two negative log-likelihood (NLL) loss functions. Note that each training data provides only a single stochastically sampled value, thus putting AgentNet for VM in the difficult condition of trying to identify the general decision rule with only one sample for each environment.

As a result, AgentNet for VM achieved a NLL loss of $-1.365$ for the test data, while the theoretically computed NLL loss was $-1.524$. We note that other approaches, such as naive MLPs, failed to achieve meaningful prediction and resulted in a NLL loss of around $+1.0$ for the VM. Figure \ref{Fig3}A visualizes the $x$-variable attention weight $\alpha^x$ of two sample agents, $a_1$ and $a_2$. AgentNet for VM accurately learned the interaction boundary of the given VM, which resembles a major sector of the circle. As Fig. \ref{Fig3}B shows, the fully trained AgentNet assigns a high value to its $x$ and $y$-variable attention only for neighbor agents, while the untrained AgentNet has no distinction between neighbor and outside cells. The predicted position distributions for these two sample agents are depicted in Fig. \ref{Fig3}C. We observe that AgentNet precisely estimated the ground-truth distribution with true mean, even though the given training data is sampled from a stochastic distribution and did not match the expected mean value. This shows the capability of AgentNet to learn the general transition rule governing the entire set, rather than merely memorizing every single training datapoint and overfitting them. We also report that AgentNet shows the same outcome with unseen test data.

\begin{figure*}
  \centering
  \includegraphics[width=\linewidth]{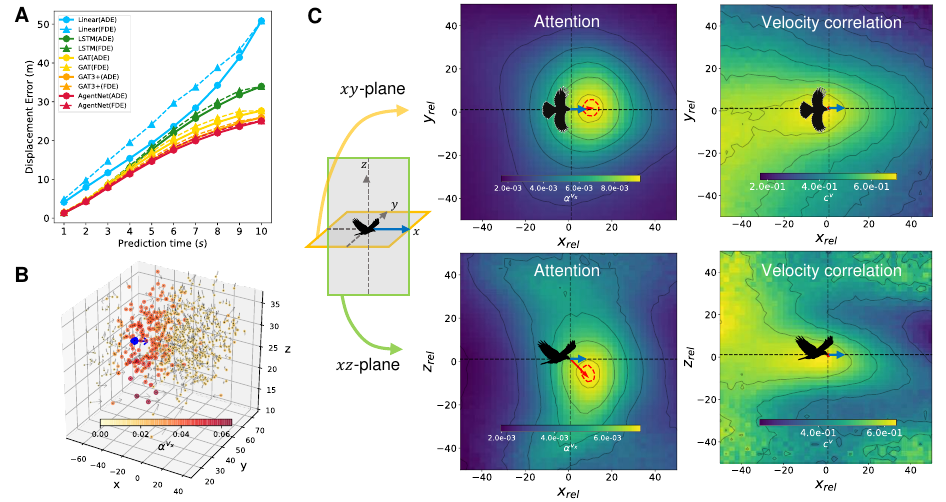}
  \caption{Result of AgentNet for CS. (A) Displacement errors of linear extrapolation, naive LSTM, GAT, GAT3+, and AgentNet. AgentNet shows the lowest displacement error compared to the baselines. Here, the final displacement error (FDE) of step $n$ indicates the averaged error of birds for which their trajectories terminated at step $n$. All of the results are averaged value from three trials. (B) Exemplary snapshot of the visualized attention of a single agent (blue circle) from the test dataset. (C) Two-dimensional heatmap of averaged attention $\alpha^{v_x}$ and cosine similarity of the velocity with respect to the relative coordinates. We align every bird's heading direction in the test dataset to the $x$-axis (blue arrow) and draw cross-sections in the $xy$-plane (upper panels) and $xz$-plane (lower panels). Different from the velocity correlation, attention shows more concentrated and strongly directional distributions that coincide with previous literature about the bird's visual frustum and sight direction. For attention, a contour of the top 0.01\% of the attention value is visualized (red, dashed) as well as the direction of the maximum attention value (red arrows).}
  \label{Fig5}
\end{figure*}

\subsection{Active Ornstein--Uhlenbeck Particle}

Differing from the Vicsek model, some active matter shows a time-correlation of particle positions due to the force inherent in the particles that allows them to move. These systems are generally referred to as self-propelled particles, which can be described by overdamped Langevin equations for the position $\mathbf{x_i}$ of each particle as
\begin{equation}
  \mathbf{\dot{x}}_i = \mu(\mathbf{F}^{\text{ext}}_i + \mathbf{F}^{\text{int}}_i) + \sqrt{2\gamma T}\boldsymbol{\eta}_i + \mathbf{f}_i ,
\end{equation}

where $\mu$ is the mobility of the particle, $\gamma$ the drag coefficient, and $T$ is temperature. Here, $\mathbf{F}^{\text{ext}}_i$ is the external potential, and $\mathbf{F}^{\text{int}}_i = -\nabla_i V$ is the total force exerted on particle $i$ due to the soft-core potential from other particles, $V = \exp(-|r_{ij}|^3/R^3)$, that depends on relative distance $r_{ij}$ and interaction length $R$. In this study, we use AOUPs confined in a harmonic potential as an example system, describing the intrinsic propulsion force $f_i$ as an independent Ornstein--Uhlenbeck process as
\begin{equation}
  \tau\mathbf{\dot{f}}_i = -\mathbf{f}_i + \sqrt{2D_a}\mathbf{w}_i,
\end{equation}

where $\tau$ is correlation time, $D_a$ is a diffusion constant, and $\mathbf{w}_i$ is a standard Gaussian white noise. As an external potential, we apply a weak harmonic potential $\mathbf{F}^{\text{ext}}_i = -k\mathbf{x}_i$ with spring constant $k=0.1$ to confine the particles, as broadly assumed and experimentally employed \cite{AOUP_harmonic}. This model is known to exhibit a collective clustering phenomenon, with the periodicity of the resulting hexagonal pattern known to be approximately $1.4R$ with no $\mathbf{F}^{\text{ext}}$ \cite{AOUP_hexagonal}.

AgentNet for AOUP adopts an LSTM model as an encoder to enable iterative data generation. The model observes 8 steps of trajectories as input data, and the loss is calculated for the next 12 steps. The model receives four state variables, $\mathbf{x}^t, \mathbf{y}^t, \mathbf{v_x}^t,$ and $\mathbf{v_y}^t$, and global variable $R$ ranging from $2.0$ to $4.0$, and predicts the parameters for four 1D Gaussian distributions, similar to the AgentNet for VM (see Appendix B for details). Note that the internal variable, $\mathbf{f}_i$, which has its own Ornstein--Uhlenbeck dynamics, is not present in the input data and thus the neural network \textit{has to infer} this hidden variable by eight steps of past trajectory. 

First, we compare the average displacement error (ADE) and final displacement error (FDE) of our model among 12 predicted steps as in previous works \cite{DM_socialgan, DM_sociallstm} along with a linear extrapolation and naive LSTM without the graph attention core as baselines. AgentNet for AOUP showed ADE/FDE of $0.041/0.064$, while linear extrapolation and LSTM showed much lower performances of $0.210/0.465$ and $0.158/0.316$, respectively. The performance of our model also exceeds the modern architectures like GAT3+ (GAT with 3-headed attention and transformer architecture), which showed the performance of $0.065/0.087$. Figure \ref{Fig4} summarizes the result of AgentNet for AOUP. Our model precisely predicted the future trajectories subject to the past states, as depicted in Fig. \ref{Fig4}A where 100 trajectories sampled from the ground-truth Langevin equation and AgentNet for AOUP are drawn. Figure \ref{Fig4}B shows that AgentNet is also capable of predicting the untrained region of the global variable $R$ and further exhibits a collective behavior that occurs far beyond the trained time scale. Since our model can iteratively predict future states indefinitely, we tested our model to predict a total of $42$ steps, which is $30$ more steps than the model was originally trained for. Surprisingly, our model predicts a precise hexagonal pattern of periodicity $7$, which coincides with the theoretical value of periodicity when $R = 5$. This verifies a generalization capability since the model had never been trained in the $R = 5$ condition and yet still properly captured the collective phenomenon, which only occurs at a much longer timescale than its training data had.

Moreover, we demonstrate that the attention $\alpha^{q}$ corresponds to the internal force $F^{\text{int}, q}$, up to a constant factor, as we claimed in system formulation section. Fig. \ref{Fig4}C and \ref{Fig4}D verifies this by showing the attention for $x$ and $y$-directional velocity $v_x, v_y$ and the magnitude of corresponding internal force $F^{\text{int}, x}, F^{\text{int}^y}$, which clearly exhibits a strong linear relationship. This cannot be achieved by a single-valued attention from conventional GAT, which shows a poor agreement with any of the force components. We report that the single attention value from GAT tries to convey the sum or average of each interaction strength. In Fig. 4E, we draw the scaled attention for $v_x$ and the internal force of the $x$ direction $F^{\text{int}, x} = \nabla V_{ij} = (-3r_{ij}^2\exp{[-r_{ij}^3/R^3]})/R^3$ versus the relative distance to the target particle $r_{ij}$. Despite a slight disagreement at small $r_{ij}$, scaled attention with constant factor $c = 0.28$ well matches $F^{\text{int}, x}$ and therefore can be considered as a good approximation for interaction magnitude. (See Appendix D for further investigation on AOUP attention.) AgentNet for AOUP successfully predicted and investigated one of the most complex systems possessing internal potential, external potential, memory effects, and stochastic noises. We note that variables other than $v_x$ also showed similar linear relationships with corresponding forces (results for other variables are reported in \cite{github}.) 

\subsection{Chimney swift trajectory}

Finally, we demonstrate the capability of our framework by predicting the empirical trajectories of a freely behaving flock of chimney swifts (CSs). Bird flocks are renowned for their rich diversity of flocking dynamics, for which models with various mechanisms such as velocity alignment and cohesion have been proposed in the last several decades \cite{VC_origin, CS_statbird, CS_vf1}. We employed here a portion of the data from \cite{CS_dryad}, recorded in Raleigh, North Carolina, in 2014. Since half of the trajectories last less than $150\text{f} = 5\text{s}$ and $80$\% last less than $300\text{f} = 10\text{s}$ due to occlusion and the limited sight of the camera, observation data takes the form of a spatiotemporal graph with dynamic nodes where each agent lasts a short period and then disappears. Thus, discarding non-full trajectories as in previous works \cite{DM_sociallstm, DM_socialgan} would significantly reduce the number of birds to consider at a given time step. To handle these disjointed yet entangled pieces of trajectories, we propose a novel inspection method that examines the data at every step of the LSTM to manually connect the hidden states from the past, exclude the nonexistent birds at a certain time, and start a new chain of hidden states from a separate neural network if an agent newly enters the scene. While several previous approaches could handle graphs with dynamic edges \cite{AN_dynamic1, AN_dynamic2, AN_dynamic3}, AgentNet is, to the best of our knowledge, the first attempt to deal with dynamic nodes on a spatiotemporal graph (see Appendix A for a formal explanation of the inspection scheme).

The number of total birds appearing in each set varied from $300$ to $1800$, and each trajectory in the set started and ended at different times. The model received state variables that exist at the current time step, produced statistics of three-dimensional position and velocity, and then the sampled states were fed back into the model for the next time prediction. NLL losses were calculated at every LSTM step for existing birds. 

Figure 5 summarizes the results of AgentNet for CS. The predictive power of AgentNet is illustrated in Fig. 5A, where linear extrapolation and naive LSTM show mostly similar results while AgentNet shows greatly reduced errors at predicting longer time steps, achieves better performances than GAT and GAT3+. Figure 5B, showing the visualized attention of a typical bird, clearly indicates the near-sighted and forward-oriented nature of the bird's interaction range. To further verify this interaction range, we averaged the attention values from the first step of predictions according to the relative coordinates of the target bird. The averaged results for $\alpha^{v_x}$ from the entire test set are drawn in Fig. 5C along with the averaged cosine similarity of the velocity, which is a commonly used measure to find the range of interaction. The interaction range projected on the $xy$-plane coincides with previous literature about biological agents' visual frustum, which depends on forward-oriented sight and the relative distance from each agent \cite{TC_chimney, DM_fish, CS_vf1, CS_vf2}. Also, the bird's $z$-directional attention is relatively concentrated \textit{downwards}; this predicted attention is physiologically plausible since downward-oriented visual fields are widely reported in various types of birds due to their foraging nature and the blind area from the beak \cite{CS_vf_xz1, CS_vf_xz2}. 

Interestingly, Fig. 5C shows that the velocity correlation on the $xy$-plane and $xz$-plane shows no particular directional tendency as attention does. Although many studies employ state correlations between agents to figure out the characteristics of interaction \cite{CS_colbehav3, CS_colbehav4}, correlation might be significantly different from the interaction range itself \cite{CS_colbehav2}. Different from correlations, our model provides a \textit{causal} interaction strength since the attention value is strongly connected to the predictability of future dynamics, which is quite useful for inferring and modeling the microdynamics of individual birds. 

Our model with variable-wise attention can further verify important physical insights. For instance, we have found that although the scale is different, the form of attention concentration is surprisingly the same regardless of the directions (Results for other variables are reported in \cite{github}). This directional homogeneity strongly implies that the bird-bird interaction is more like a near-sighted version of the Vicsek model, differs from the distance-based force models like AOUP which must exhibit directional heterogeneity. In conclusion, AgentNet employed the position and velocity (heading direction) of neighboring birds into its prediction, thereby showing better prediction compared to the non-interactive baseline and qualitatively plausible interaction range.

\section{Conclusion}

This study proposed AgentNet, a generalized framework for the data-driven modeling of a complex system. We demonstrated the flexibility, capability, and interpretability of our framework with large-scale data from various complex systems. Our framework is \textit{universally} applicable to agent-based systems that are governed by pairwise interactions and for which a sufficient amount of data is available. The proposed framework can infer and visualize variable-wise interaction strength between agents, which could assist researchers in gaining clearer insights into given systems and their dynamics. Furthermore, AgentNet is scalable for an arbitrary number of agents due to the nature of GNNs, thus facilitating free-form simulation of the desired system with any initial condition. Since attention values from our model can be directly interpretable as a variable-wise interaction strength function, we expect that AgentNet will be useful in heterogeneous settings where each state variable interacts with different neighbors.

There are a great number of domains in which AgentNet is anticipated to exhibit its full potential. As we demonstrated via AOUP and CS, the analysis of active matter such as bacterial cells \cite{AOUP_harmonic, CC_bacteria}, animal flocks \cite{DM_flock1, DM_flock2, DM_flock3, CS_colbehav1, CS_statbird}, or pedestrian dynamics \cite{DM_sociallstm, DM_socialgan} may greatly benefit from our approach.  Also, since GNNs were originally proposed for data with graph structures, AgentNet may yield data-driven models of both agent and node dynamics of a network by incorporating an adjacency matrix instead of assuming a complete graph. AgentNet can retrieve the underlying graph and interaction strength from data, which encompasses the research fields of epidemic dynamics \cite{CC_epidemics}, network identification, and various inverse Ising problems \cite{DM_inverse1}. We could further apply different encoders and decoders to improve the performance and include available domain knowledge. For instance, a Gaussian mixture model \cite{TC_vrnn} or variational model \cite{DM_socialgan} that could approximate an arbitrary distribution may be suitable to approximate multimodal or highly irregular distributions.

One limitation of the current work is that AgentNet cannot fully capture three or higher orders of interactions. The pairwise assumption is nearly the only inductive bias we have imposed on our model, which will require modification if the target system is expected to have strong higher-order interaction. First, by increasing the number of message passing layers, GNNs can employ information from further than one-hop neighbors and possibly capture the higher-order interactions among three or more agents. Another way to alleviate the pairwise assumption is to consider higher-order interactions directly in network construction. Applying a GNN with a hypergraph structure \cite{PD_hypergraph1, PD_hypergraph2, PD_hypergraph3}, one of the rapidly growing research areas in machine learning, to AgentNet would be a direct extension of the current study.

We highlight the virtually unbounded scope of the proposed framework in this study, and hope that AgentNet shines a new light on physical modeling and helps researchers in diverse domains delve into their systems in a data-driven manner.

\begin{acknowledgements}
  This research was supported by the Basic Science Research Program through the National Research Foundation of Korea NRF-2017R1A2B3006930. We appreciate Y.J. Baek for providing insights to model demonstration and fruitful discussion.
\end{acknowledgements}

\appendix
\section{Neural architecture of AgentNet and training details}

\subsection{AgentNet and baseline implementations}
We implemented our AgentNet model with \texttt{PyTorch} \cite{EX_pytorch}. The encoder and decoder layers of AgentNet are composed of multi-layer perceptrons (MLPs). The dimension notation such as [32, 16, 1] means that the model consists of three perceptron layers with $32$, $16$, and $1$ neurons in each layer. Also, dims. is an abbreviation of dimensions.

All of the encoding layers of AgentNet are composed of \textbf{[Input dims, 256, Attention dims]}. Here, input dimensions are chosen as the sum of the number of state variables and additional variables, such as global variables (as in AgentNet for AOUP) or indicator variables (as in AgentNet for CS). The form of the final dimension indicates that each output of the encoder (key, query, and value) will be processed separately. See \cite{AN_trans} for more details about transformer architecture.

With these outputs and (additional) global external variables ($\boldsymbol{u}$), neural attention is applied to calculate attention value $\alpha_{ij}$ from encoded data $e(\boldsymbol{s^t})$. First, the algorithm constructs $a_{ij}^q$, the attention coefficient for the $q$-th state variable between agents $i$ and $j$, and feeds the concatenated vectors into MLP(att) as

\begin{equation}
  a_{ij}^q = \text{Att}(\text{Key}(e(s^i)), \text{Query}(e(s^j)), \boldsymbol{u})
\end{equation}

and applies the sigmoid function

\begin{equation}
  \alpha_{ij}^q = \frac{1}{1 +{\exp(-a_{ij}^q})}
\end{equation}

where $\text{Key}$, $\text{Query}$, and $\text{Att}$ indicate the corresponding MLPs used for transformer architecture and has dimensions.

After variable-wise attentions are multiplied to their respective values and averaged, we concatenate the (original target agent's) value and its averaged attention-weighted values (from others) and feed it into the variable-wise separated decoder. Since two tensors are concatenated, the last dimension of this tensor has twice the length of the original dimension of the value tensor. The decoder consists of \textbf{[2 $\times$ value dims., 128, output dims.]}.

In the stochastic setting (VM, AOUP, and CS), the decoded tensor further feeds into other layers to obtain sufficient statistics for the probabilistic distribution. In this paper, those statistics are means and variances of state variables. MLP layers for these values consist of \textbf{[output dims., 64, corresponding number of variables]}. For instance, AgentNet for CS has $2 \times 6 = 12$ separate layers to calculate means and variances for $6$ state variables.

In the case of a target system with probable time correlations, we adopted long short-term memory (LSTM) as an encoder to capture the correlations \cite{AN_LSTM}. Hidden states and cell states have $128$ dims. each and are initialized by additional MLPs that are jointly trained with the main module. As explained in the main manuscript, AgentNet checks at each time step whether an agent is new and present. When an agent is newly entered, new LSTM hidden states are initialized. Otherwise, hidden states succeed from the previous result.

For the baseline, we employed a MLP, LSTM, and GAT model where the variable-wise graph attention module is missing. For MLP and LSTM, We doubled the number of layers and neurons of the decoder to compensate for the missing attention module, which its decoder consists of \textbf{[2 $\times$ value dims., 256, 256, output dims.]}. For standard GAT \cite{IN_gat}, we left everything the same as AgentNet and replaced variable-wise attention core to original graph-attention core with linear projection matrices of \textbf{[Attention dims., 128]} and inner-product attention was used with those 128-dimensional vectors. For GAT3+, we implemented multi-headed attention (with 3 heads) with dimensions of $12$ (for AOUP) and $32$ (for CS), and every other module is the same as AgentNet. This choice is to (Note that even for GAT and GAT3+, we used \textbf{[Input dims, 256, Attention dims]} dimensions of encoding layers for the key, query and value, instead of linear projection matrices as the original architecture.)

\begin{figure}
  \centering
  \includegraphics[width=\linewidth]{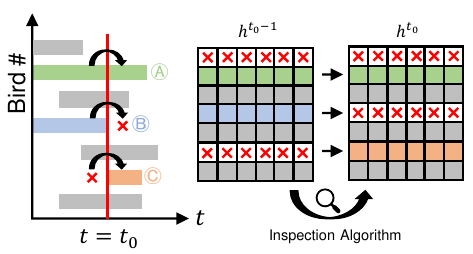}
  \caption{Trajectory data schema of the chimney swift flock \cite{TC_chimney}. (left) In each time step, each agent (bird) could take one of three states: (A) it continues to exist, (B) it disappears, or (C) it newly enters the scene. (right) The hidden states of the agents for LSTM change by the inspection algorithm, correctly removing vanished agents and introducing new agents with initialized hidden states. Red crosses indicate dummy variables (zeros) for padding.}
  \label{Fig6}
\end{figure}

\subsection{Inspection scheme of AgentNet for CS}

As the author of the original paper noted \cite{TC_chimney}, CS data contains a lot of short trajectories due to frequent occlusion and limitation of the camera viewing angle. An exemplary snapshot of typical trajectory data is drawn in Fig. 6, where all of the trajectories are not aligned in time and have different lengths. This form of unstructured data can be expressed as a dynamic graph that precludes the naive application of a graph neural network (GNN) \cite{IN_gcn} with LSTM  \cite{AN_LSTM} states since the neighbor of each node changes dynamically and each node has a different starting time. Hence, we need an inspection method to manually check the hidden state at each time step and update each node's status to the appropriate form.

Figure \ref{Fig6} describes all three possible cases that our inspection scheme needs to deal with. Case A is an ordinary case with a continued state, so we can simply hand over the previously updated hidden state to the next iteration. Case B means that the agent, which existed at the last time step ($t = t_0-1$), disappeared and should not be considered as a valid neighbor anymore. Our inspection method excludes such data and creates a mask for the attention matrix to ensure that the attention between a valid agent and non-existent agent should be strictly zero at any time. Finally, case C indicates a newly entered agent whose hidden states should be initialized before it starts its chain of hidden states.

In practice, input data is created from raw data in the form of \textbf{[time step, number of total agents, number of state variables]}. Note that we use the number of \textit{total} agents for every time step, which is not always the same as the number of agents at the specific time. For instance, in Fig. \ref{Fig6}, there are 7 birds present in the total scene, but only $6$ birds are present at $t = t_0$. Instead of changing the data size every time step, we set every dimension to the maximum agent number (in this case, $7$) and fill the currently non-existing birds' state variables with dummy values (we used zeros). With such padding, we can create a batched data set that enables parallel calculations to speed up the GPU deep learning. But we have to carefully mask them out properly at each forward pass to ensure that none of these dummy data affects the results because we keep calculating and updating these dummy data as well as the real agents' data.

Our inspection algorithm creates two masks with length of total agents, which is called 'Now mask' $m_{now}$ and 'New mask' $m_{new}$. Each mask checks the existence of agents at the current($t = t_0$) and next time step ($t = t_0+1$), and assigns $1$ if the corresponding condition is satisfied and $0$ otherwise. 'Current mask' assigns $1$ if the agent is present at $t_0$. 'New mask' assigns $1$ if the agent is not present at $t_0$ but appears in $t_0+1$. Inspection method employs these three masks to control hidden state updates, attention calculations, and state variables updates.

At the start of every iteration, the algorithm checks new agents that need to be initialized. Let us denote the hidden states of agents from time $t$ as $h^t$ and state input data as $x^t$. Then, the hidden state update can be expressed as

\begin{equation}
  h^{t+1} = h^t \odot (1-m_{new}) + I(x^t) \odot m_{new} \label{eq:a1}
\end{equation}

where $I(x)$ indicates the hidden state initialization module consisting of neural layers, and $\odot$ means element-wise multiplication. Then, after the attention calculation, the algorithm casts a two-dimensional attention mask $m_{att}$ on raw attention matrix $\alpha^q_{raw}$ to exclude all of the attention values between real and dummy agents as follows:

\begin{equation}
  m_{att, ij} = \begin{cases}
    1 & \text{if} \ a_i \ \text{and} \ a_j \ \text{are both real agent} \\
    0 & \text{if otherwise}
  \end{cases}\label{eq:a2}
\end{equation}

and $\alpha^q = \alpha^q_{raw} \odot m_{att}$. We can construct $m_{att}$ from the 'Now mask' by repeating row vector mask $m_{now}$ for the column dimension to make expanded matrix mask $M_{now}$, with $m_{att} = M_{now} \times M_{now}^T$.

\subsection{Training scheme}
All training used $2$ to $10$ NVIDIA TITAN V GPUs, with which the longest training for a single model took less than two days. Mish activation function \cite{SI_mish} with a form of $f(x) = x\text{tanh}(\text{softplus}(x))$ and the Adam Optimizer \cite{AN_adam} were used for the construction of models and training. The learning rate was set to 0.0005 and decreased to 70\% of the previous value when the test loss remained steady for 30 epochs. In the case of AgentNet for CS, we employed weighted NLL loss for different time steps, in which weights are inversely proportional to the frequency of the sample with a given trajectory length, to resolve the imbalance of available trajectory length. Table III shows further details of the model for each system, including the number of attention heads.

\begin{table*}\centering
  \caption{Implementation details of the models for sample systems. $*$ : global variable ($R$), $**$ : indicator variable}
  \begin{ruledtabular}
    \begin{tabular}{lcccc}
      System                                 & Input data dims. & Output data dims. & Attention head dims. & Layer composition    \\ \hline
      1. Cellular Automata                   & 3                & 1                 & 16                   & [32, 32, 16, 1]      \\
      2. Vicsek Model                        & 4                & 2                 & 16                   & [32, 64, 32, 1]      \\
      3. Active Ornstein--Uhlenbeck Particle & 4                & 4                 & 16                   & [32+1$^*$, 16, 8, 1] \\
      4. Chimney Swift                       & 6 + 1$^{**}$     & 6                 & 96                 & [192, 16, 8, 1]
      \label{table:3}
    \end{tabular}
  \end{ruledtabular}
\end{table*}

\section{Dataset Details}
\subsection{Cellular automata}
We employed the cellular automata (CA) model based on the rule of Conway's life game \cite{CA_conway}. As described in the main manuscript, CA takes place on regular grids with each cell on the grid altering its cell state at each time step according to a specific set of rules. These rules are often notated as $B3/S23$, which means a dead cell regenerates with three neighboring live cells, while a live cell stays alive with two or three neighboring live cells. In this study, we used a $ 14 \times 14$ grid with uniformly random initial cell states and updated an inner square grid of $12 \times 12$ to avoid the periodic boundary problem. We simulated 1,000 sets of samples for demonstration, comprising 800 samples for training and 200 samples for testing. We report that fewer samples such as 500 or 300 also resulted in a perfectly trained model with 100\% test accuracy. Each target cell along with its eight neighbor cells yields a total of $2^9 = 512$ possible microstates. We located a $3 \times 3$ microstate template at a random position on the grid cell, initialized to random cell states, and produced AgentNet output. If the model correctly learned the transition rule, it would result in theoretical output assigned by the transition rule of CA regardless of its position and other irrelevant cell states.

From the perspective of AgentNet, CA is a binary classification problem, i.e. whether each cell becomes alive or dead at the next time step. AgentNet for CA receives three state variables: positions $\mathbf{x}^t$ and $\mathbf{y}^t$, and cell state $\mathbf{c}^t$.The output here is a list of expected probabilities that each cell becomes alive. We use the binary cross entropy loss function between the AgentNet output and the ground-truth label as
\begin{equation}
  L_{BCE}(\mathbf{s}) = \sum_{i} y_i\log{c_i} + (1-y_i)\log{(1-c_i)},
\end{equation}

where $c_i$ is the predicted $i$th cell state of output $\mathbf{s}$, and $y_i$ is the corresponding true label of the $i$th cell.

\subsection{Vicsek model}
The Vicsek model (VM) \cite{VC_origin} assumes that flocking occurs due to velocity alignment with neighbors. Among the many variants, we implemented the simplest model with alignment terms and positional Gaussian noise. (Note that this differs from the originally proposed model \cite{VC_origin}, which used angular Gaussian noise instead.) In VM, the $i$th agent interacts with the $j$th agent if the distance between the two agents, $r(s_i^t, s_j^t)$, is smaller than a certain range, $r_c$, and the absolute value of the angle between the heading direction of the $i$th agent and the position of the $j$th agent, $\theta(s_i^t, s_j^t)$, is smaller than a certain angle, $\theta_c$. This interaction range models the sight range limit of living organisms, resulting in a circular sector form. In VM, the $i$th agent averages the velocity among its interacting neighbors $R_i$ and adds Gaussian noise $\mathcal{N}(0, \sigma)$ to compute its velocity $\mathbf{v}_i^{t+1}$, described as follows:

\begin{figure}[h]
  \begin{align*}
    \mathbf{v}_i^{t+1} = (\mathbf{v}_i^{t} + \sum_{a_j \in R_i} \mathbf{v}_j^{t})/(|R_i|+1) + \mathcal{N}(0, \sigma).
  \end{align*}
\end{figure}

This formula is equivalent to $(\sum_{R^*_i} \mathbf{v}_i)/|R^*_i| + \mathcal{N}(0, \sigma)$ where $R^*_i = R_i \cup {a_i}$.

AgentNet for VM aims to predict the position of the next time step, which is a sum of the current position and calculated velocity. We simulated 2,000 sets of samples for demonstration, comprising 1,600 samples for training and 400 samples for testing. In our simulation, $r_c = 1$, $\theta_c = 120^{\circ}$, and standard deviation of noise $\sigma = 0.2$.

The model infers the parameters of two 1D Gaussian distributions, which is means $(\mu^x, \mu^y)$ and $\sigma^x, \sigma^y$. We calcualte the sum of the NLL loss function for Gaussian distribution to train the AgentNet for VM as
\begin{equation}
  L_{NLL}(\mathbf{s}) = \sum_{q} \sum_{i} -\frac{1}{2}\log(2\sigma_i^q) + \frac{(\mathbf{y}_i^q - \mu_i^q)^2}{2(\sigma_i^q)^2}, \label{eqn:VM1}
\end{equation}

where $\mu_i^q$ and $\sigma_i^q$ are the predicted statistics of variable $q$ of the $i$th agent, and $\mathbf{y}_i^q$ is the corresponding label of the $i$th cell. In the VM case, $q = \{x, y\}$.

\subsection{Active Ornstein--Uhlenbeck particle}

For the AOUP dataset, we simulated 8,000 sets of training data and label pairs of 100 particles. All particles were uniformly spread on a circle of radius $5$, and initial speeds were sampled from uniform distribution $\mathcal{U}(0, 0.05)$.
We implemented the Euler--Maruyama \cite{SI_EM} method with timestep $dt=0.01$ to numerically simulate the AOUP trajectories. The data and label points were further subsampled from the simulated trajectory with a frequency of $10$ Hz, which means every 1 out of 10 subsequent data points were chosen. In terms of real-time, our model receives 0.8 s of data observation time and then predicts the following 1.2 s of trajectory. The following constants were adopted for simulation: $\gamma = 1$, $\tau = 0.5$, $k = 0.1$, $D_a = 0.02$, and $T = 0.2$. We have found that our model is robust against change in system constants, showing similar performance with different constant values.

We attached a constant vector $R$ to the attention vector to open the possibility that interaction function $h_{pair}$ depends on the global variable $R$. (In case of AOUPs, this is the case since $R$ affects interaction potential $\mathbf{F}^{int}$.) Differing from the VM case, AgentNet for AOUP yields a total of 8 parameters for each agent, namely means and variances for positions $\mathbf{x}^{t+1}$, $\mathbf{y}^{t+1}$ and velocities $\mathbf{v_x}^{t+1}$, $\mathbf{v_y}^{t+1}$. This is necessary because input and output state variables should be the same in order to iteratively sample from the predicted distribution and feed it into the prediction at the next step.

We employed teacher forcing \cite{SI_teacherforcing} to train the LSTM-based AgentNet, which is a technique that feeds ground-truth labels into subsequent LSTM cells instead of sampled output in the early stages of training. This is useful to stabilize the training of trajectory prediction since the prediction depends on the last output which typically explodes to meaningless values in early, untrained stages. We set an initial epoch of 50 as the teaching period such that the possibility of using the ground-truth label is $1-\text{epoch}/50$. This gradually decreasing possibility becomes $0$ at epoch 50, after which ground-truth is not used.

For evaluation, linear extrapolation along with naive LSTM are selected as baselines. In the case of extrapolation, $x$ and $y$ coordinates of the previous 8 steps are extrapolated through time and the next 12 consecutive steps are recorded. The naive LSTM used the same state variables ($x, y, v_x, v_y$) and the same output structure, but only contained the LSTM encoder and MLP decoder (i.e. was missing an attention core). This implies that no effects of interactions with others are considered by the naive LSTM. Averaged displacement error (ADE) is calculated by taking the average of Euclidean distances from ground-truth to predicted coordinates for all 12 steps. Final displacement error (FDE) only takes the averages of the final (12th step) error.

\subsection{Chimney swift flock trajectory}

The original paper \cite{TC_chimney} aimed to focus on collective behavior during the landing sequence of a chimney swift (CS) flock. The flock data contains 30 min of observed CS trajectories at 30 frames (=$\text{f}$) per second, with approximately 100,000 unique trajectories and a maximum of 1,848 birds at one instance. The reason that the number of unique trajectories greatly exceeds the size of the flock is because many of the trajectories from the same birds are treated separately if (1) they escape from the sight of the cameras and later re-enter, or (2) the birds are occluded by other birds thus introducing ambiguity. The whole dataset can be divided into three parts: (1) an initial stage where the birds are starting to gather, (2) an intermediate stage where the flock forms with birds showing collective spinning, and (3) the final stage of landing on the chimney. In this study, we used the first portion (file A) of the data since we are interested in general bird flocking rather than a specific landing sequence (file C), and the second part (file B) contained more than 3000 unique birds in a $5$ s instance, generally exceeding the memory capacity of a GPU ($12$ GB) even for the case of a single data per batch. Among 30 min (= 54000 f) of trajectory data, we employed around 10 min (= 18000 f) of frames and constructed a dataset with $30\text{f} * 10 = 300\text{f}$ each, $20\text{f}$ apart from each other. Although a single datapoint spans an overlapped time range with other data, we strictly split the training and test dataset to remove any possibility of data contamination. Exact details about the dataset and statistics can be found in \cite{TC_chimney}.

In the CS case, there is a problem with unbalanced labels since not every sample has full 10-step (10 s) trajectories, as mentioned in the main manuscript. Thus, we checked the number of agents present at certain steps and calculated a weighted loss to strengthen the effect of cases with fewer birds (typically, the case of a higher time step has a smaller number of constituents since many trajectories end early).

AgentNet for CS was trained with 530 sets of bird trajectories that spanned 300 frames each, split into 10 time steps. Since every trajectory starts and ends at different time steps, we take an average of bird displacement error where its starting point is regarded as time step 0. For instance, if the trajectory of bird 1 spans from time step 0 to 3 and the one of bird 2 spans from 4 to 6, the error calculated at bird 1, step 2 and bird 2, step 6 will be treated the same as the displacement error at  the time step \textit{2} since both steps are 2nd steps to their starting point. Also, since there were trajectories only lasting 2 time steps ($=1.0 \text{s}$), we cannot apply linear regression for those trajectories. Instead, we linearly extrapolated the trajectories by employing the given velocity at the first time step. Also, we intentionally scaled the input data by multiplying $0.1$ for every variable to stabilize the training while preserving the relative difference of variable magnitude.

\begin{figure}%[tbhp]
  \centering
  \includegraphics[width=\linewidth]{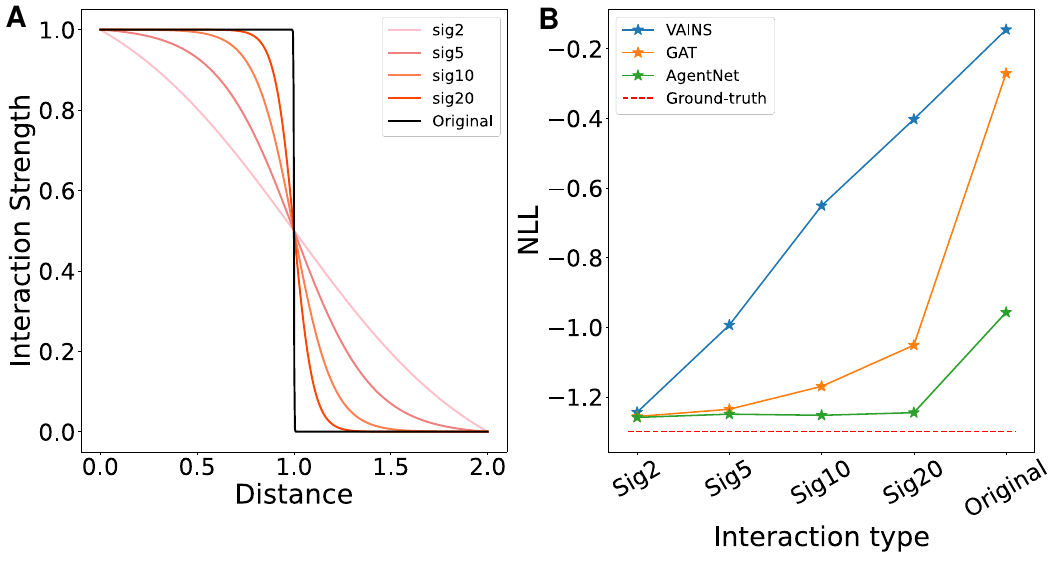}
  \caption{(A) Functions of a \textit{smoothed} version of interaction range from sigmoid models with various smoothing parameter $b$. (B) Performance comparison of VAINS, GAT, and AgentNet compared to NLL loss from groundtruth.} \label{Fig7}
\end{figure}

\begin{figure}
  \centering
  \includegraphics[width=\linewidth]{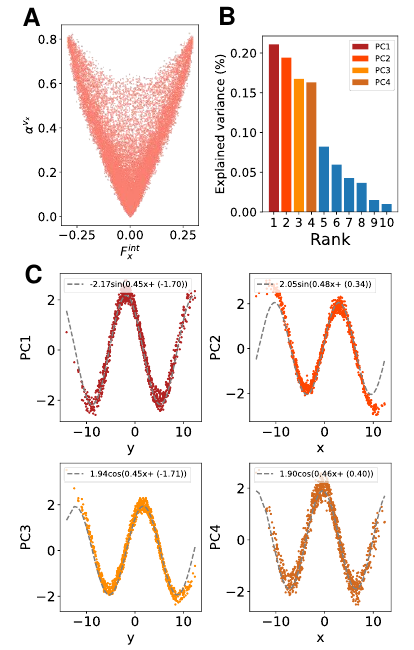}
  \caption{PCA result of value vectors $v$ from AgentNet for AOUP. (A) $x$-directional force $F^{int,x}$ abd attention for $v_x$. Note that attention learns magnitude not the sign, due ot its non-negative constraint. (B) Explained variance of each principal component sorted by rank, which shows four clearly distinguished components. (B) Each principal component and corresponding $x$ or $y$ coordinate data.}
  \label{Fig8}
\end{figure}

\section{Advantages of Neural Attention}

Neural attention requires a greater number of parameters to optimize, but showed greater performance in predicting complex agents compared to existing attention mechanisms, as shown in the main manuscript. One possible reason for the outperformance of neural attention is that the interaction range of a complex system is far more nonlinear and complicated than conventional attention mechanisms can handle, which are relatively under-parametrized for such interaction ranges. This assumption is further supported if the performance gap decreases as the interaction boundary becomes more linear. We experimented with a \textit{smoothed} version VM (Fig. 7A), where interaction strength is defined as a sigmoid function $s(x) = 1/(1+\exp[b(x-a)])$. Figure 7B shows the performances of VAIN (exponential-based attention from \cite{IN_vain}), GAT (conventional transformer), and AgentNet (neural attention) with different $b$ when $a = 1$. Note that the original interaction boundary coincides with the $b \rightarrow \infty $ case. Results illustrate the difference between the scales of the two mechanisms for $b$, which underpins the aforementioned nonlinearity hypothesis.

\section{AOUP attention analysis}

The variable-wise attention in AgentNet is wrapped by the sigmoid function, where its output is equal to or greater than $0$, as seen in Fig. \ref{Fig7}A. Hence, relative positional information is vital to convert the magnitude into actual force with a sign, which can be directly applied to change the velocity. For instance, force should be $+x$ direction if the target agent $i$ is in $(1,0)$ and the neighbor agent $j$ is in $(0,0)$, but it changes to $-x$ direction if the neighbor agent moves to $(2,0)$, although the magnitude (and thus the attention value) would be the same. It is possible to regain this directional information if the value vector $v_j$, corresponding to a leftover function, contains the positional data of the $j$th agent; the position information of the $i$th agent can be delivered by $h_{\text{self}}$.

Figure \ref{Fig7} verifies that value vector $v$ preserves the positional information of the input data. We gathered data from 100 test samples of AOUP, and performed principal component analysis (PCA) to 16-dimensional vector $v$ from the encoder. Figure \ref{Fig7}A shows that the top four principal components (PCs) stand out in terms of explained variance, compared to the PCs with lower ranks. We found that these four PCs have an interesting nonlinear relationship with two coordinates $x$ and $y$ from input data. Figure \ref{Fig7}B shows that these four PCs have strong sinusoidal relationships with $x$ and $y$, giving clear evidence of information preservation. We note that this specific non-linear transformation is not special, and there could be various other possible means of preserving positional information. (For instance, the simplest way to convey positional information would be to directly preserve the coordinate data from the input without any nonlinear transformation.) In this particular case, we can assume that the overall function $f$ could successfully retrieve the original coordinate from these sinusoidal functions and ultimately obtain the sign of the relative position by comparing it with the target $i$th agent's position.

%\bibliographystyle{apsrev}
%\bibliography{prx}% Produces the bibliography via BibTeX.


\end{document}